\documentclass[11pt]{article}

\usepackage{fullpage}
\usepackage{amssymb}
\usepackage{amsmath}
\usepackage{graphicx}
\usepackage{boxedminipage}
\usepackage{xy}
\xyoption{all}

\newtheorem{theorem}{\bf Theorem}[section]

\newtheorem{lemma}[theorem]{\bf Lemma}

\newtheorem{proposition}[theorem]{\bf Proposition}
\newtheorem{claim}[theorem]{\bf Claim}

\newcommand {\myproof} {{\it Proof: }}

\newcommand {\qed} {\hfill$\Box$}

\newcommand{\eat}[1] {}
\newcommand{\calA} {{\cal A}}
\newcommand{\calB} {{\cal B}}
\newcommand{\calC} {{\cal C}}
\newcommand{\calP} {{\cal P}}
\newcommand{\calQ} {{\cal Q}}
\newcommand{\calI} {{\cal I}}
\newcommand{\calS} {{\cal S}}

\newcommand{\calY} {{\cal Y}}
\newcommand{\calX} {{\cal X}}
\newcommand{\calR} {{\cal R}}
\newcommand{\wh}[1] {\widehat{#1}}
\newcommand{\wt}[1] {\widetilde{#1}}
\newcommand{\pair}[2] {\langle #1, #2\rangle}

\title{Distributed and Parallel Algorithms for Set Cover Problems with Small Neighborhood Covers}
\author{Archita Agarwal \and Venkatesan T. Chakaravarthy \and Anamitra R. Choudhury \and Sambuddha Roy \and Yogish Sabharwal}
\date{IBM Research Lab, New Delhi, India\\
  \texttt{\{archiaga,vechakra,anamchou,sambuddha,ysabharwal\}@in.ibm.com}
}
\begin{document}

\maketitle

\begin{abstract}
In this paper, we study a class of set cover problems that satisfy a special property which 
we call the {\em small neighborhood cover} property.
This class encompasses several well-studied problems including vertex cover, interval cover, 
bag interval cover and tree cover. We design unified distributed and parallel algorithms that can handle 
any set cover problem falling under the above framework and yield constant factor approximations.
These algorithms run in polylogarithmic communication rounds in the distributed setting
and are in NC, in the parallel setting.
\end{abstract}

\section{Introduction}
In the classical set cover problem, we are given a set system $\pair{E}{\calS}$,
where $E$ is a {\em universe} consisting of $m$ {\em elements}
and $\calS$ is a collection of $n$ subsets of $E$.
Each set $S\in \calS$ has cost $w(S)$ associated with it.
The goal is to select a collection of sets $\calR\subseteq \calS$
having the minimum aggregate cost such that every element is included in at least one
of the sets found in $\calR$.

There are two well-known classes of approximation algorithms for the set cover problem \cite{ShmoysBook}.
The first class of algorithms have an approximation ratio of $O(\log \Delta)$,
where $\Delta$ is the maximum cardinality of the sets in $\calS$.
The second class of algorithms have an approximation ratio of $f$,
where $f$ is the {\em frequency parameter} which is the maximum number of sets of $\calS$
that any element belongs to. 
The above approximation ratios are nearly optimal \cite{Hypergraph-Hardness, RazS97, Feige}. 
In general the parameters $\Delta$ and $f$ can be arbitrary and so the above algorithms do not yield
constant factor approximations.
The goal of this paper is to develop parallel/distributed constant factor approximation algorithms
for certain special cases of the problem.

In the parallel setting, we shall use the NC model of computation and its randomized version RNC.
Under this model, Rajagopalan and Vazirani \cite{RV} presented a randomized 
parallel $O(\log m)$-approximation algorithm for the general set cover problem.
Under the same model, Khuller et al. \cite{KVY} presented a $(f+\epsilon)$-approximation algorithm
for any constant frequency parameter $f$ and $\epsilon>0$.

In the distributed setting, we shall adopt a natural communication model which has also been used in prior work.
In this model, there is a processor for every element 
and there is a communication link between any two elements $e_1$ and $e_2$,
if and only if both $e_1$ and $e_2$ belong to some common set $S\in \calS$.
We shall view the element itself as the processor.
Each element has a unique ID and knows all the sets to which it belongs.
We shall assume the standard synchronous, message passing model.
The algorithm proceeds in multiple communication rounds, where in each round
an element can send a message to each of its neighbors in the communication network.
We allow each element to perform a polynomial amount of processing in each round
and the messages to be of polynomial size.
We are interested in two performance measures: 
(i) the approximation ratio achieved by the algorithm; and
(ii) the number of communication rounds. 
Ideally a distributed algorithm should have polylogarithmic communication rounds.
Under the above distributed model, Kuhn et al. \cite{near-sighted} and 
Koufogiannakis and Young \cite{Kouf-Young} presented distributed algorithms for the general set cover problem
with approximation ratios of $O(\log \Delta)$ and $f$, respectively;
both the algorithms run in polylogarithmic communication rounds.

There are special cases of the set cover problem wherein both $\Delta$ and $f$ are arbitrary,
which nevertheless admit constant factor approximation algorithms.
In this paper, we study one such class of problems satisfying a criteria that we call the  
{\em small neighborhood cover} property (SNC-property).
This class encompasses several well-studied problems such as 
vertex cover, interval cover and tree cover.
Furthermore, the class subsumes set cover problems with a constant frequency parameter $f$.
Our results generalize the known constant factor approximation algorithm for the latter class.

Our goal is to design unified distributed and parallel algorithms
that can handle any set cover problem falling under the above framework.
In order to provide an intuition of the SNC-property,
we next present an informal (and slightly imprecise) description of the property.
We then illustrate the concept using some example problems and intuitively show why these
problems fall under the framework.
The body of the paper will present the precise definition of SNC set systems. 

{\bf SNC Property.}
Fix an integer constant $\tau\geq 1$. We say that two elements are neighbors, if some $S\in \calS$
contains both of them. The neighborhood of an element is defined to be the set of all its neighbors (including itself).
We say that an element $e\in E$ is a $\tau$-SNC element,
if there exist at most $\tau$ sets that cover the neighborhood of $e$.
The given set system is said to have the $\tau$-SNC property, 
if for any subset $X\subseteq E$, the set system restricted\footnote{the restricted set system is $\pair{X}{\calS'}$, 
where $\calS'=\{S\cap X~:~S\in \calS\}$} to $X$ contains a $\tau$-SNC element.
The requirement that every restriction has a $\tau$-SNC element will be useful in solving the problem iteratively.

{\bf Example Problems.}
We next present some example $\tau$-SNC set cover problems.

{\em Vertex Cover: }Given a graph $G$, we can construct a set system
by taking the edges as the elements and the vertices as sets.
In this setup, an element belongs to only two sets and hence, the set systems defined by the
vertex cover problem satisfy the $2$-SNC property.
In general, set cover problems having a constant frequency parameter $f$ would induce
$\tau$-SNC set systems with $\tau=f$.

\begin{figure}
\begin{center}
\includegraphics[width=5in]{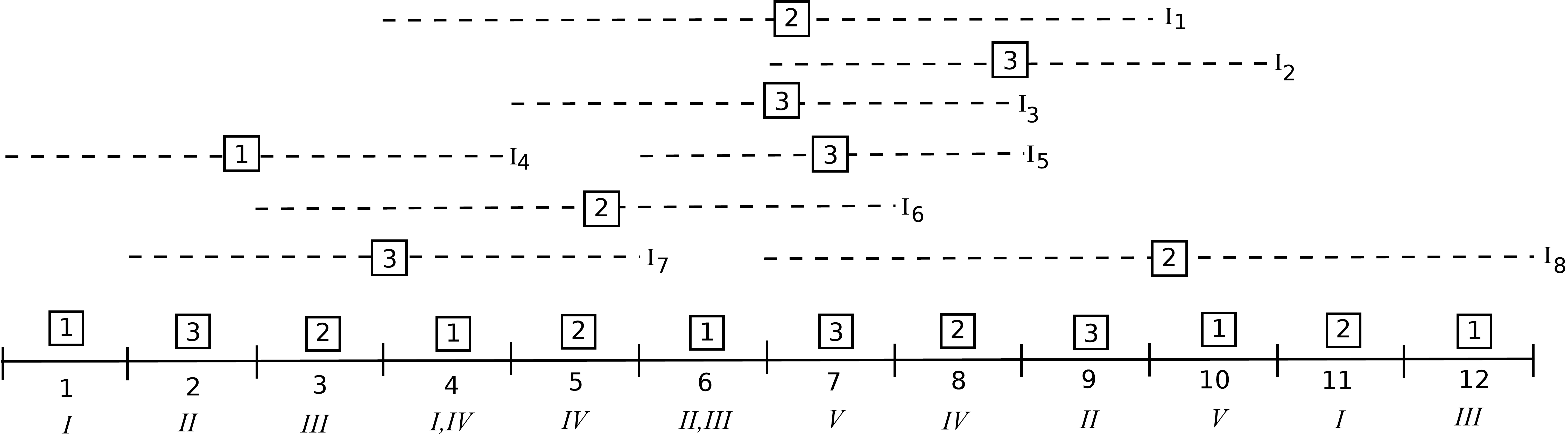}
\caption{Illustration for interval cover problems}
\label{fig:interval}
\end{center}
\end{figure}

{\em Interval Cover: }
In this problem, we are given a timeline divided into some $m$ discrete timeslots
$1,2,\ldots,m$. The input includes a set of intervals $\calI$, where each interval $I\in \calI$
is specified by a range $[s(I),e(I)]$, where $s(I)$ and $e(I)$ are the starting and ending points of $I$.
Each interval $I$ also has an associated cost $w(I)$. We say that an interval $I$ covers a timeslot
$1\leq e\leq m$, if $e\in [s(I),e(I)]$. The goal is to find a collection of intervals
having minimum aggregate cost such that every timeslot $t$ is covered by at least one interval in the collection.
We can view the problem as a set cover instance by taking the timeslots to be the elements and
taking each interval $I\in \calI$ as a set consisting of the timeslots covered by $I$. 
See the picture on the left in Figure \ref{fig:interval} for an illustration (ignore the Roman numerals).
Consider any timeslot $e$ and let $\calQ\subseteq \calI$ be the set of intervals covering $e$.
Among the intervals in $\calQ$, the interval $I_l$ with the minimum starting point
and the interval $I_r$ having the maximum ending point can cover the neighborhood of $e$ (resolving ties arbitrarily).
For example, for timeslot $3$, $I_l = I_4$ and $I_r = I_1$.
Hence, the set systems defined by the interval cover problem satisfy the $2$-SNC property.

\begin{figure}
\begin{center}
\includegraphics[width=3in]{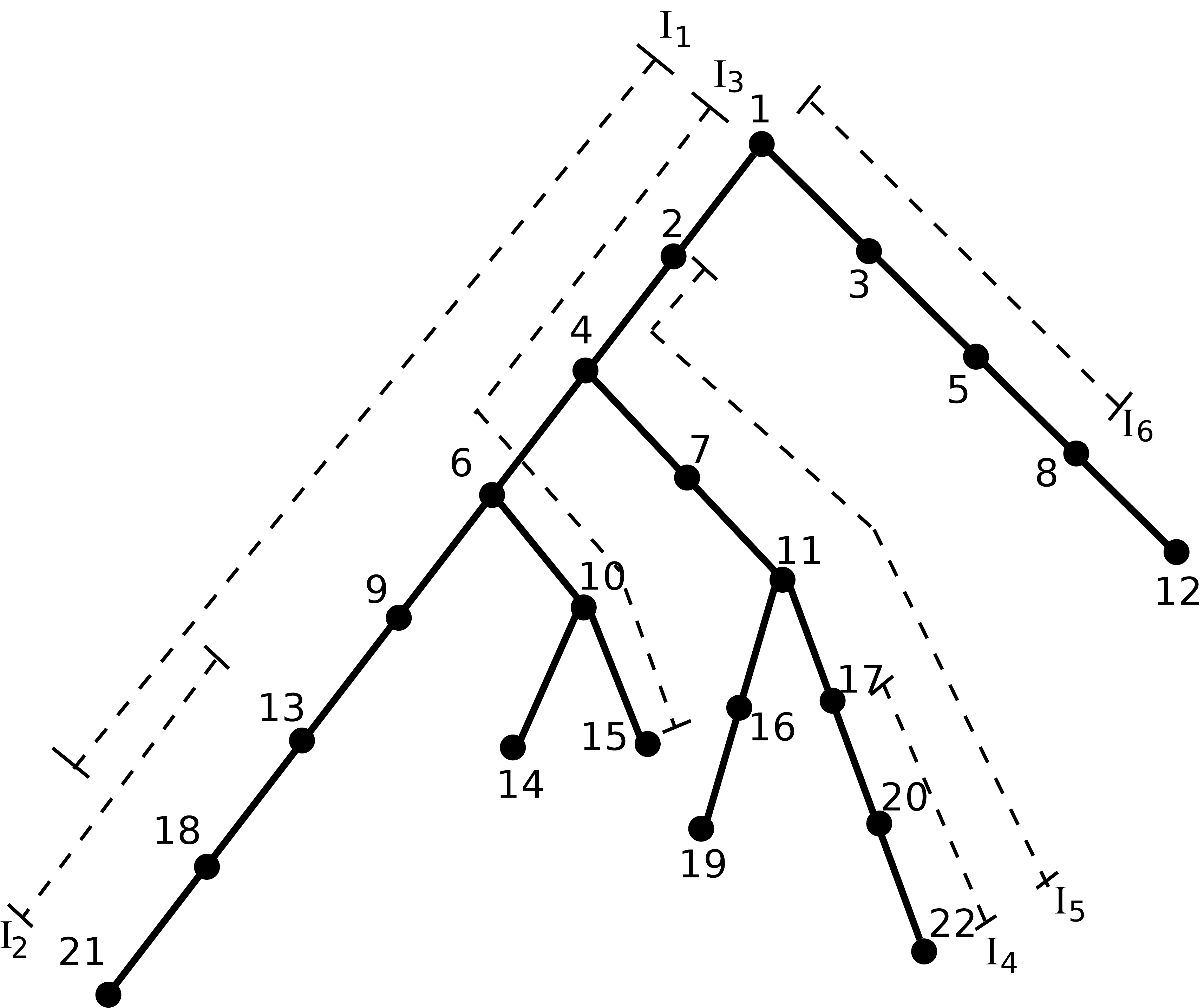}
\caption{Illustration for tree cover}
\label{fig:tree}
\end{center}
\end{figure}

{\em Tree Cover Problem: }
In the tree cover problem, we are given a {\em rooted} tree $T=(V,H)$.
The input includes a set of intervals $\calI$, where each interval is specified as a pair of nodes $\pair{u}{v}$
such that $u$ is an ancestor of $v$. The interval $I$ can be visualized as the path from $u$ to $v$.
The interval is said to cover an edge $e\in H$, if $e$ is found along the above path.
Each interval $I$ has a cost $w(I)$ associated with it.
The goal is to find a collection of intervals of minimum cost covering all the edges.
We can view the problem as a set cover instance by taking the edges to the elements
and taking the intervals as sets.
It is not difficult to see that the tree cover problem generalizes the interval cover problem. 
See the picture on the right in Figure \ref{fig:tree} for an illustration.
Consider any leaf edge $e$. Let $\calQ$ be a set of intervals covering the edge $e$.
Among the intervals in $\calQ$, let $\wh{I}$ be the interval extending the most
towards to the root. Note that $\wh{I}$ covers the neighborhood of $e$.
For example, in the figure, for the leaf edge $\pair{20}{22}$, the interval $I_5$ will serve as $\wh{I}$.
Thus, any leaf edge satisfies the $1$-SNC property.
It is not difficult to see that any restriction will also contain an element satisfying the $1$-SNC property. 
Hence, the set systems defined by the tree cover problem satisfy the $1$-SNC property.

{\em Bag Interval Cover Problem: }
This problem generalizes both vertex cover and interval cover problems.
The input consists of a timeline divided into discrete timeslots $\{1,2,\ldots, T\}$.
We have a set of $n$ intervals $\calI$. 
Each interval $I \in \calI$ has a starting timeslot $s(I)$, an ending timeslot $e(I)$ and a weight $w(I)$. 
Timeslots are grouped into $m$ {\em bags} $B_1, B_2, \ldots, B_m$; a timeslot may belong to more than one bag.
The interval $I$ is said to cover a bag $B_i$, if it spans at least one timeslot from the bag $B_i$.
The goal is to find a collection of intervals having minimum aggregate cost such that
each bag is covered by some interval in the collection. 
The {\em girth} of the system is defined to be the maximum cardinality of any bag and it is denoted $g$; 
Viewed as a set cover problem, each bag will correspond to an element and 
each interval will correspond to a set. 
See the picture on the left in Figure \ref{fig:interval} for an illustration. The bag number are shown in Roman numerals.
For instance, Bag I consists of timeslots $\{1,4,8\}$. The girth of the system is $3$.

Consider any element (bag) $B$ containing timeslots $\{e_1, e_2, \ldots, e_r\}$ (with $r\leq g$).
For each timeslot $e_i$, among the intervals spanning $e_i$ select the intervals
having the minimum staring point and the maximum ending point.
This set of $2r$ intervals can cover the neighborhood of $B$.
Thus any element satisfies the $2g$-SNC property.
Hence, the set systems defined by the bag interval cover problem satisfies the $2g$-SNC property.

{\em Priority Interval Cover: }
As in the case of interval covering,
we are given a discrete timeline $[1,m]$ and a set of intervals $\calI$.
In addition, each timeslot $e$ has a priority $\chi(e)$ (a positive integer)
and similarly, each interval $I\in \calI$ is also associated with a priority $\chi(I)$.
An interval $I$ can cover a timeslot $e$, if $e\in [s(I),e(I)]$ and $\chi(I)\geq \chi(e)$.
The basic interval covering problem corresponds to the case where there is only one priority.
Let the number of priorities used be $K$. See Figure \ref{fig:interval} for an illustration;
the numbers within boxes show the priorities of intervals and timeslots.
The interval $I_4$ cannot cover timeslot $2$, even though the interval spans the timeslot.

Consider any timeslot $e$ having the highest priority. As in the interval cover problem,
among the set $\calQ$ of intervals covering $e$, the intervals having the minimum starting point
and the maximum ending point put together can cover the neighborhood of $e$.
Thus all timeslots $e$ having the highest priority would be $2$-SNC elements. 
It is not difficult to argue that any restriction will also contain an element satisfying the $2$-SNC property. 
Hence, the set systems defined by the priority intercal cover problem satisfy the $2$-SNC property.

{\bf Layer Decomposition.}
An important concept that will determine the running time of our algorithms is that of layer decomposition.
We present an intuitive description of layer decomposition. 
The formal definition will be presented in the body of the paper.

Consider a set system $\pair{E}{\calS}$ satisfying the $\tau$-SNC property for some constant $\tau$.
Let $Z_1$ be the set of all $\tau$-SNC elements in the given set system.
Let $Z_2$ be the set of $\tau$-SNC elements in the set system obtained by restricting to $E-Z_1$.
Proceeding this way, for $k\geq 2$, let $Z_k$ be the set of $\tau$-SNC elements in the set system 
obtained by restricting to $E-(Z_1\cup Z_2\cup \cdots Z_{k-1})$.
We continue the process until no more elements are left.
Let $L$ be the number of iterations taken by this process.
The sequence $Z_1, Z_2, \ldots, Z_L$ is called the {\em layer decomposition} of the set system $\pair{E}{\calS}$.
Each set $Z_k$ is called a {\em layer}. The number of layers $L$ is called the {\em decomposition length}.
The decomposition length of the input set system is of importance, since the running time of our parallel/distributed
algorithms depend on this quantity.

We next study the decomposition length for our example problems.
In the case of vertex cover, interval cover and bag interval cover problems,
we saw that all the elements satisfy the $\tau$-SNC property in the given system $\pair{E}{\calS}$ itself.
Hence, the decomposition length of these set systems is one.
In the tree cover problem, recall that all the leaf edges in the given tree $T$ are $1$-SNC elements.
Thus, all the leaf edges will belong to the first layer $Z_1$. Once these leaf edges are removed,
the leaf edges in the remaining tree will belong to the second layer $Z_2$.
Proceeding this way, we will get a layer decomposition in which the number of layers will be
the same as the depth of the tree; later, we describe how to reduce the decomposition length to be $O(\log m)$.

In this paper, we will only focus on set cover problems having logarithmic decomposition length 
and derive distributed/parallel algorithms with polylogarithmic rounds/running-time for such problems.
We note that there are set cover problems that induce $\tau$-SNC systems with a constant $\tau$, 
but having arbitrary decomposition length. An example for the phenomenon is provided by the priority interval cover problem. 
In this case, all the timeslots having the highest priority would belong to layer $Z_1$.
In general, the timeslots having priority $k$ will belong to layer of index at most $K-k+1$, where $K$ is the total number of priorities.
Therefore the number of layers would be the could be as high as the the number of priorities.

{\bf Our Results. }In this paper, we introduce the concept of $\tau$-SNC property.
We note that all the example problems considered earlier can be solved optimally or within constant factors
using the primal-dual paradigm. All these algorithms have certain common ingredients;
these are abstracted by $\tau$-SNC framework.
We present three algorithms for the set cover problem on $\tau$-SNC set systems.
\begin{itemize}
\item
A simple sequential $\tau$-approximation algorithm.
\item 
A distributed $\tau$-approximation algorithm for $\tau$-SNC set systems of logarithmic decomposition length.
The algorithm is randomized and uses $O(\log^2 m)$ communication rounds.
\item
A parallel $(1+8\tau^2)$-approximation algorithm for $\tau$-SNC set systems of logarithmic decomposition length.
The algorithm can be implemented in NC.
\end{itemize}

Our algorithms have the following salient features:
\begin{itemize}
\item
They provide unified constant factor approximations for set cover problems falling under the $\tau$-SNC 
framework with logarithmic decomposition length, in both distributed and parallel settings.
\item
A surprising and interesting characteristic of these algorithms is that they are model independent. 
Meaning, they only require the set system as input and do not need the underlying model defining the set system.
For instance, in the tree cover problem, the algorithms do no need the structure of the tree as input.
At a technical level, we show that the layer decomposition can be constructed by considering only the local 
neighborhood information; this fact is crucial in a distributed setting.
\end{itemize}

Regarding the example problems, we saw that in case of the vertex cover, interval cover and bag-interval cover problems,
the decomposition length is one. Thus our parallel and distributed algorithms will apply to these problems.
The case of tree cover problem is more interesting.
As we observed earlier, the set systems arising from the tree cover problem are $1$-SNC set systems,
however the the decomposition length is the same as the depth of the tree, 
which could be as large as $\Omega(m)$ (where $m$ is the number of edges). 
Hence our parallel and distributed algorithms cannot be applied to this case.
However, we shall show that it is possible to reduce the decomposition length to $O(\log m)$, 
if we settle for a slightly higher SNC parameter of $\tau=2$:
\begin{itemize}
\item
We prove that the set systems
defined by the tree cover problem satisfy the $2$-SNC property with decomposition length $O(\log m)$.
\end{itemize}
In other words, the tree cover problem instances induce a $1$-SNC set systems
of arbitrary decomposition length, as well as $2$-SNC set systems of decomposition length $O(\log m)$.
Using the above fact, we can apply our parallel and distributed algorithms and obtain constant factor apporoximations.

It is easy to see that for any constant $f$, set systems with frequency parameter $f$
satisfy the $\tau$-SNC property, with $\tau=f$.
Dinur et al. \cite{Hypergraph-Hardness} proved that for any $f\geq 3$, 
it is NP-hard to approximate the set cover problem within a factor of $(f-1-\epsilon)$, for any $\epsilon>0$.
Thus, the approximation ratio of the sequential and distributed algorithms are nearly optimal.
In the parallel setting, we present an algorithm with an approximation ratio of $(1+8\tau^2)$.
Improving the approximation ratio is an interesting open problem.

While this is the first paper to consider the general $\tau$-SNC framework, the specific example
problems have been studied in the sequential, parallel and distributed settings.
Improved algorithms are known in specific cases.
We next present a brief survey of such prior work and provide a comparison to our results.

{\bf Comparison to Prior Work on Example Problems. }
For the vertex cover problem, sequential $2$-approximation algorithms are well known \cite{ShmoysBook}.
In the parallel setting, Khuller et al. \cite{KVY} presented a parallel NC algorithm
having approximation ratio of $2+\epsilon$, for any $\epsilon>0$ (see also \cite{Gandhi}).
Koufogiannakis and Young \cite{Kouf-Young} presented the first parallel algorithm with approximation ratio of $2$.
Their algorithm is randomized and runs in RNC.
The above algorithms can also be implemented in the distributed setting
(see also \cite{GKP}).

The interval cover problem can be solved optimally in the sequential setting via dynamic programming.
Bertossi \cite{Bertossi} presented an optimal parallel (NC) algorithm,
which can also handle the more general case of circular arc covering.
However, their algorithm requires the underlying model (i.e., the timeline and intervals) explicitly as input.
We are not familiar with prior work on the problem in the distributed setting.

Chakrabarty et al. \cite{CGK} study the tree cover problem and its generalizations under the sequential setting.
In this setting, the problem can be solved optimally via dynamic programming or the primal-dual paradigm.
Furthermore, the constraint matrices defined by the problem are totally unimodular (see \cite{CGK}).
We are not familiar with any prior work on parallel/distributed algorithms for this problem.
For this problem $\tau=2$ and so, our sequential/distributed algorithms provide an approximation ratio of $2$.
The parallel algorithm has an approximation factor of $33$. However, we note that one of the reasons for the high
ratio is that the algorithm is oblivious to the underlying model.

The priority interval cover problem is studied by Chakrabarty et al. \cite{CGK} and Chakaravarthy et al. \cite{our-esa}. 
They provide polynomial time optimal algorithms based on the dynamic programming.
To the best of our knowledge, the bag interval cover problem has not been considered before.
However, the notion of bag constraints has been considered in the related context
of interval packing problems (see \cite{Bar-Noy-Jacm, Berman-Dasgupta}).
Covering integer programs (CIP) generalize the set cover problem.
These are well studied in both sequential and distributed settings (see \cite{Kouf-Young,CGK}, and references therein).

{\bf Proof Techniques. }
All the algorithms in the paper utilize the primal-dual paradigm. 
The sequential algorithm is fairly straightforward and it is similar to that of
the primal-dual algorithm $f$-approximation algorithm for the set cover problem.
The latter algorithm works by constructing a maximal feasible solution to the dual
which would automatically yield an $f$-approximate integral primal solution.
Our problem requires two additional ingredients. The first is that an arbitraty maximal dual
solution would not suffice. Instead, the solution needs be constructed in accordance
with the layered decomposition. Secondly, a maximal dual solution would not automatically
yield a $\tau$-approximate integral primal solution. A reverse delete phase is also needed.
In this context, we present a polynomial time algorithm for computing the layer decomposition of the given set system,
which can also be implemented in both parallel and distributed settings.

In the distributed setting, the only issue is that the above steps need to be performed within 
polylogarithmic number of rounds. We address the issue by grouping the elements
based on the Linial-Saks decomposition \cite{LS} of the communication network.

The parallel algorithm is more involved and forms the main technical component of the paper.
For a general set system, Khuller et al. \cite{KVY} (see also \cite{Gandhi}) 
present a parallel procedure for computing nearly maximal dual solution with maximality parameter of $(1-\epsilon)$,
using the idea of raising several dual variables simultaneously.
However, the parallel running of the procedure is $O(f\log(1/\epsilon)\log m)$, where $f$ is the frequency parameter.
In our problems, the parameter $f$ could be arbitrary and the above running time is not satisfactory.
We present a procedure that produces a near maximal solution with maximality parameter $1/8$.
While the maximality parameter is worse compared to prior work, the running time of our procedure is independent of $f$.
This procedure could be of independent interest. The procedure is similar in spirit to that of Khuller et al.,
but the analysis for bounding the number of iteration takes a different approach.

As mentioned earlier, our setting requires an additional reverse delete phase,
whose parallelization poses interesting technical issues.
Our procedure executes the phase by processing the layer decomposition in a zig-zag manner.
In iteration $i$, the procedure processes layer $i$ and performs the reverse delete for the particular layer.
However, this involves revisiting the older layers $1, 2, \ldots, i-1$.
Each step involves computing the maximal independent set of a suitable graph, for which we utilize
the parallel algorithm due to Luby \cite{Luby}.
The overall number of steps would be $O(L^2)$ (where $L$ is the decomposition length)
and the approximation ratio is $8\tau^2$ (as against the ratio $\tau$ achieved by the sequential/distributed algorithms).

Our algorithm raises two interesting technical problems.
The first is that whether we can construct a near maximal solution to the dual with parameter $(1-\epsilon)$,
while keeping the parallel running time independent of the frequency parameter $f$.
Secondly, whether the reverse delete can be performed in parallel while achieving a primal complementary
slackness parameter of $\tau$. An affirmative answer to either question would result in improved approximation algorithms.

\section{Preliminaries}
In this section, we present the formal definition of the $\tau$-SNC property and related concepts.
We also present algorithms for computing the layer decomposition for a given $\tau$-SNC set system.

{\bf $\tau$-SNC Element: }Fix an integer constant $\tau\geq 1$. Consider a subset of elements $X\subseteq E$ and an element $e\in X$.
Let $\calQ\subseteq \calS$ be the collection of all sets that contain $e$.
The element $e$ is said to be a {\em $\tau$-SNC element within $X$}, 
if for any $\calP \subseteq \calQ$,
there exist at most $r$ sets $S_1, S_2, \ldots, S_r\in \calP$ (with $r\leq \tau$)
such that every element in $e\in X$ covered by $\calP$ is also covered by one of the $\tau$ sets:
\[
\bigcup_{S\in \calP} S\cap X = \bigcup_{i=1}^r S_i\cap X.
\]
Note that the $\tau$ sets must be selected from the collection $\calP$.
The property is trivially true if $|\calP|\leq \tau$, but it becomes interesting if $|\calP|\geq \tau+1$.

{\bf $\tau$-SNC Set System: }The given set system $\pair{E}{\calS}$ is said to be a {\em $\tau$-SNC set system}
if for every subset of elements $X\subseteq E$, there exists an element $e\in X$
which is a $\tau$-SNC element within $X$. 
The set system is said to be a {\em total $\tau$-SNC set system},
if for every subset $X\subseteq E$, every $e\in X$ is a $\tau$-SNC element within $X$.
The following property is easy to verify.

\begin{proposition}
\label{prop:AAA}
If an element $e\in X$ is a $\tau$-SNC element within $X$, then for any $Y\subseteq X$ such that $e\in Y$,
$e$ is also a $\tau$-SNC element within $Y$.
\end{proposition}

However, the converse of the above statement may not be true.
Namely, an element $e$ may be a $\tau$-SNC element within a set $X$, but it may not be a $\tau$-SNC element
within a superset $Y\supset X$. To see this, suppose $\calP$ is a collection of sets 
such that every $S\in \calP$ contains $e$.
The collection $\calP$ may cover an element $x\in Y-X$,
which may not be covered by some $\tau$ sets of $\calP$ that cover the neighborhood of $e$ within $X$.

{\bf Layer Decomposition: }
Consider a $\tau$-SNC set system $\pair{E}{\calS}$.
The notion of layer decomposition is defined via an iterative process, as described in the introduction.
Let $Z_1$ be the set of $\tau$-SNC elements within $E$.
For $k\geq 2$, let $Z_k$ be the set of $\tau$-SNC elements within $E-(Z_1\cup Z_2\cup \cdots \cup Z_{k-1})$
We terminate the process when there are no elements left. Let $L$ be the number of iterations taken by the process.
The sequence $Z_1, Z_2, \ldots, Z_L$ is called the {\em layer decomposition} of the given set system.
Each set $Z_i$ is called a layer and $L$ is called the {\em decomposition length}
We consider $Z_1$ to be the left-most layer and $Z_L$ as the right-most layer.

{\bf Computing Layer Decompositions: }
As part of our algorithms, we will need a procedure for computing the layer decomposition of a 
given $\tau$-SNC set system. The following lemma provides such a procedure. 
The proof is given in Section \ref{sec:layer-compute}

\begin{lemma}
\label{lem:TTT}
There exists a procedure for computing the layer decomposition of a given $\tau$-SNC set system.
In the sequential setting, it can be implemented in polynomial time.
In the distributed setting, it can be implemented in $O(L)$ communication rounds.
In the parallel setting, the algorithm takes $L$ iterations each of which can be implemented in NC.
\end{lemma}

{\it Remark: }Notice that any $\tau_1$-SNC set system is also a $\tau_2$-SNC set system
for any $\tau_2\geq \tau_1$. The decomposition length of the system will depend on the choice of $\tau$.
The procedure stated in the lemma will produce the layer decomposition corresponding to the value of $\tau$ provided 
as input to the procedure.

\section{Sequential Algorithm}
In this section, we present a sequential $\tau$-approximation algorithm for solving the set cover problem
restricted to $\tau$-SNC set systems, for a constant $\tau$.
The parallel and distributed algorithms build on the sequential algorithm.
As mentioned in the introduction, our example problems
can be solved optimally or approximately using the primal-dual paradigm. 
All these algorithm have certain common ingredients in the design and analysis,
which are captured by the notion of $\tau$-SNC property.
Our algorithm for the general $\tau$-SNC set systems 
also goes via the primal-dual paradigm and utilizes ideas from the algorithms for the example problems.
The pseudocode for the algorithm is given in Figure \ref{fig:seq-pseudo}.

The primal and the dual for the input set system $\pair{E}{\calS}$ are given below.

\begin{tabular}{p{3in}p{3in}}
\begin{eqnarray*}
\min \quad \sum_{S \in \calS} x(S) \cdot w(S) \\
\sum_{S\in \calS ~:~ e \in S} x(S)  \geq  1 \quad  (\forall e\in E) 
\end{eqnarray*}
&
\begin{eqnarray*}
\max \quad \sum_{e \in E} \alpha(e) \\
\sum_{e\in S} \alpha(e)  \leq  w(S) \quad (\forall S\in \calS)
\end{eqnarray*}
\end{tabular}

The primal LP includes a variable $x(S)$ for each set $S\in \calS$
and a constraint for each element $e\in E$.
The dual includes a variable $\alpha(e)$ for each element $e\in E$ (corresponding to the primal constraint)
and a constraint for each set $S\in \calS$ (corresponding to the primal variable).
The primal and the dual would also include the non-negativity constraints $x(S)\geq 0$ and $\alpha(e)\geq 0$.

Let the input set system be $\pair{E}{\calS}$ having $m$ elements and $n$ sets.
Using the procedure given in Lemma \ref{lem:TTT}, compute the layer decomposition $Z_1, Z_2, \ldots, Z_L$.
Obtain an ordering $\sigma$ of the elements by placing the elements in $Z_1$ first,
then those in $Z_2$ next and so on; the elements in $Z_L$ will appear at the end of the ordering
(within a layer, the elements can be arranged arbitrarily).
Let $\sigma=e_1, e_2, \ldots, e_m$ be the ordering produced by this process.
Notice that for $k\geq 1$, the element $e_k$ is a $\tau$-SNC element within $\{e_k, e_{k+1}, \ldots, e_m\}$.
The $\tau$-approximation algorithm would exploit the above ordering.

The algorithm works in two phases: a forward phase and a reverse-delete phase.
The forward phase would produce a dual feasible solution $\wh{\alpha}$ and a cover $\calA$ for the system.
In the reverse-delete phase, some sets in $\calA$ would be deleted to get the final solution $\calB$.

The forward phase is an iterative procedure which will scan the ordering $\sigma$ from left to right. 
We start by initializing $\calA\leftarrow \emptyset$
and $\alpha(e)\leftarrow 0$, for all $e\in E$. 
In iteration $k\geq 1$, we pick an element next element $e$ from the ordering $\sigma$
which is uncovered by the collection $\calA$. We raise the dual variable $\alpha(e)$ until
some dual constraint becomes tight (i.e., LHS becomes equal to the RHS). 
Let the corresponding set be $S$. We include the set $S$ in $\calA$ and proceed to the next iteration.
The process is terminated when all the elements are covered. 
Let $\wh{E}$ be the set of elements whose dual variables were raised.

In the second phase (called reverse-delete phase), 
we shall delete some sets from $\calA$ and construct a new solution $\calB$ such that the following 
complementary slackness properties are satisfied:
\begin{itemize}
\item 
{\em Dual-slackness:} For any set $S\in \calB$, the corresponding dual constraint is tight.
\item 
{\em Primal slackness:} For any element $e \in \wh{E}$, the corresponding primal constraint is approximately tight:
\begin{eqnarray}
\label{eqn:seq-tau}
\sum_{S\in \calB~:~e \in S} x(S)  \leq  \tau \quad  (\forall e\in E) 
\end{eqnarray}
\end{itemize}
Once we ensure these properties, standard weak-duality arguments can be applied
to argue that $\calB$ is a $\tau$-approximate solution.

The reverse-delete procedure is described next.
Initialize $\calB\leftarrow \calA$.
For any element $e\in \wh{E}$, the corresponding is primal constraint is approximately tight:
Let the number of elements in $\wh{E}$ be $s$. Arrange these elements
in the order in which they were raised, say $\wh{\sigma}=\wh{e}_1, \wh{e}_2, \ldots, \wh{e}_s$.
Let $\wh{S}_1, \wh{S}_2, \ldots, \wh{S}_s$ be the sets picked by the forward phase
when these variables were raised, respectively.
Consider the sequence $\wh{\sigma}$ in the reverse order, starting with $\wh{e}_s$.
The iteration $k$ works as follows. 
Let $X$ be the set of elements that were uncovered by $\calA$ in the beginning of the iteration 
in which $\wh{e}_k$ was picked. Notice that $\wh{e}_k$ is a $\tau$-SNC element within $X$.
Let $\calP\subseteq \calB$ be the collection of sets from $\calB$ which cover $\wh{e}_k$. 
The $\tau$-SNC property ensures that we can collapse $\calP$ into at most $\tau$ sets.
Meaning, we can find sets $S_1, S_2, \ldots, S_r$ (with $r\leq \tau$) such that
\[
\bigcup_{S\in \calP} (S\cap X) = \bigcup_{j=1}^r (S_j\cap X).
\]
Delete all the sets found in $\calP$ from $\calB$ and retain only the sets $S_1, S_2, \ldots, S_r$.
In doing so, we have not lost feasibility of $\calB$. To see this, first notice that the elements in $X$ still remain
covered. Regarding the elements in $E-X$, 
the sets $\wh{S}_1, \wh{S}_2, \ldots, \wh{S}_{k-1}$ covers all these elements.
One potential issue is that some of these set could be part of the sets we deleted;
however, this is not possible, since $e_k$ was selected to be an uncovered element in the corresponding iteration
of the forward phase..
We have ensure that Equation \ref{eqn:seq-tau} holds for the element $\wh{e}_k$.
Proceeding this way, at the end of the reverse-delete phase we will obtain our output solution $\calB$.

All the elements in $\calB$ satisfy primal slackness property (Equation \ref{eqn:seq-tau}). 
Regarding the dual-slackness property, we included a set $S\in \calA$ in the forward phase,
only when the corresponding dual constraint is tight. 
Furthermore, the dual variables were not modified in the reverse-delete phase
and no new set was introduced in $\calB$.
Thus, the solution $\calB$ also satisfies the primal-slackness properties.

\begin{figure}[t]
\begin{center}
\begin{boxedminipage}{\hsize}
\begin{small}
\begin{tabbing}
xx\=xx\=xx\=xx\=xxx\=xxx\=\kill
\textbf{Begin}  \\
\> // Forward Phase:\\
\> Let $\sigma=e_1, e_2, \ldots, e_m$ be the ordering of the elements according to $\tau$-SNC property.\\
\> Initialize. $\calA \leftarrow \emptyset$. For all $e\in E$, $\alpha(e)=0$.\\
\> For $k=1,2,3, \ldots$\\
\> \> Among the elements uncovered by $\calA$, let $e_k$ be the element appearing earliest in the ordering $e_k$\\
\> \> Raise the dual variable $\alpha(e_k)$ until some dual constraint becomes tight:\\
\> \> \> $\alpha(e_k) \leftarrow \max_{S:e_k\in S} w(S) - \sum_{a\in S} \alpha(a)$\\
\> \> \> Include the corresponding set $S$ in $\calA$:\\
\> \> \\
\> // Reverse Delete Phase:\\
\> $\calB \leftarrow \calA$\\
\> Let $\wh{e}_1, \wh{e}_2, \ldots, \wh{e}_s$ be the sequence of elements whose dual variables were raised.\\
\> For $k=s$ to $1$\\
\> \> Let $X$ be the elements uncovered by $\calA$ in the beginning of the $k$th iteration.\\
\> \> Let $\calP\subseteq \calB$ be the collection of sets covering $e_k$\\
\> \> Find sets $S_1, S_2, \ldots, S_r\in \calP$ (with $r\leq \tau$) such that\\
\> \> \> all the elements in $X$ covered by $\calP$ are also covered by $S_1, S_2, \ldots, S_r$\\
\> \> Delete all the sets found in $\calP$ from $\calB$, except $S_1, S_2, \ldots, S_r$\\
\> \> Output $\calB$.\\
\textbf{End}
\end{tabbing}
\end{small}
\end{boxedminipage}
\end{center}
\caption{Sequential $\tau$-approximation algorithm}
\label{fig:seq-pseudo}
\end{figure}

\section{Parallel Algorithm for $\tau$-SNC Set Systems}
\label{sec:main}
In this section, we present a parallel algorithm for the set cover problem on $\tau$-SNC set systems with logarithmic decomposition length.
The approximation ratio of the algorithm is $(1+8\tau^2)$.
Similar to the sequnatial algorithm, the parallel algorithm also proceeds in two phases, a forward phase and a reverse-delete phase.
A pseudocode for the algorithm can be found in Figure \ref{fig:par}

\subsection{Forward Phase}
Consider a pair of solutions $\pair{\calA}{\alpha}$,
where $\calA\subseteq \calS$ is a feasible cover and $\alpha$ is a dual feasible solution.
For a constant $\lambda\in [0,1]$, we say that the above pair is {\em $\lambda$-maximal},
if for any $S\in \calA$, the corresponding dual constraint is approximately tight:
\begin{eqnarray}
\label{eqn:DDD}
\sum_{e\in S} \alpha(e)  \geq  \lambda \cdot w(S)
\end{eqnarray}
In the forward phase, we shall construct a $(1/8)$-maximal solution.
The procedure runs in $O(L\cdot [\log m + \log \frac{w_{\max}}{w_{\min}}])$ iterations,
where each iteration can be implemented in NC, where $L$ is the decomposition length.
As we shall see, via a standard preprocessing trick, we can ensure that $w_{\max}/w_{\min}$ is 
bounded by $O(m)$. The process would increase the approximation ratio by an additive factor of one.
Thus when $L$ is logarithmic, the procedure runs in NC.
Furthermore, our procedure would satisfy certain additional properties to be specified later. 

{\it Remark: }
While we shall describe our algorithm for the specific scenario of $\tau$-SNC set systems,
it can handle arbitrary set systems and produce $(1/8)$-maximal solutions
in $O(\log m + \log(w_{\max}/w_{\min}))$ iterations.
The problem of finding such approximately maximal solutions in parallel for general set systems is of independent interest.
Khuller et al.\cite{KVY} (see also \cite{Gandhi}) presented procedure for computing $(1-\epsilon)$-maximal solutions, 
for any $\epsilon>0$.  Their algorithm takes $O(f\log(1/\epsilon)\log(m))$ iterations, where $f$ is the frequency parameter.
For the specific case of $f=2$ (the vertex cover scenario),
a parallel procedure for producing $1$-maximal solutions is implicit in the work of 
Koufogiannakis and Young \cite{Kouf-Young}.
Their procedure runs in $O(\log m)$ iterations.
While our procedure has inferior value on the parameter $\lambda$, the number of iteration is independent of the frequency parameter $f$.
The procedure could be independent interest.
The procedure is similar to that of Khuller et al. \cite{KVY}, but the goes via a different analysis for bounding the number of iterations.

We now discuss the forward phase.
Using the procedure given in Lemma \ref{lem:TTT}, compute the layer decomposition $Z_1, Z_2, \ldots, Z_L$, where $L$ is the decomposition length.
Initialize $\calA=\emptyset$ and set $\alpha(e)=0$, for all elements $e\in E$.
The forward phase works in $L$ epochs processing the layers from left to right.
For $1\leq k\leq L$, the goal of epoch $k$ is to ensure that 
$\calA$ covers all the elements in $Z_k$.

Consider an epoch $k$. While the goal of the previous $k-1$ epochs would have been to ensure coverage 
for $Z_1, Z_2, \ldots, Z_{k-1}$, the collection $\calA$ might already be covering some elements from $Z_k$ 
(unintentionally). Let $R_k\subseteq Z_k$ be the set of elements found in $Z_k$ which are not covered by $\calA$.
The purpose of epoch $k$ is to ensure coverage for all the elements in $R_k$.
The epoch $k$ works in multiple iterations.
Consider an iteration $j\geq 1$.
A set $S\in \calS$ is said to {\em participate} in iteration $j$, if it is not already included in $\calA$.
Similarly, an element $e\in R_k$ is said to {\em participate} in iteration $j$,
if it is not all already covered by $\calA$.
For each participating set $S$, compute: 
(i) Current degree $d_j(S)$, which is the number of participating elements found in $S$; 
(ii) Current LHS value of dual constraint of $S$: $h_j(S) = \sum_{e\in S}\alpha(e)$;
(iii) Current difference between LHS and RHS of the dual constraint of $S$:
$c_j(S) = w(S) - \sum_{e\in S} \alpha(S)$;
(iv) Current {\em penalty} for $S$: $p_j(S) = c_j(S)/d_j(S)$ 
(intuitively, if $S$ is included in $S$, $d_j(S)$ elements will be newly covered and this is the cost/penalty
each such element pays).
For each participating element $e$, compute the minimum penalty offered by each 
set covering $e$: $q_j(e) = \min_{S~:~e\in S} p_j(S)$.
Increase (or raise) the dual variable $\alpha(e)$ by $q_j(e)$.
This would raise the value of the LHS of the dual constraints.
For every participating set $S$, check if its dual constraint is approximately tight:
$\sum_{e\in S} \alpha(e)  \geq  w(S)/8.$
If the above condition is true, then add $S$ to $\calA$.
This completes the description of the iteration $j$. 
The above process is continued until all the elements in $R_k$ are covered by $\calA$.
This completes epoch $k$ and we proceed to epoch $k+1$.

Notice that any dual variable $\alpha(e)$ is raised only to an extent of its minimum penalty $q_j(e)$.
This ensures that all the dual constraints will remain satisfied at the end of each iteration.
The above procedure can be implemented in both distributed and parallel settings.
In the distributed setting, each participating element (or the corresponding node in the network)
can raise its dual variable $\alpha(e)$ independently using information obtained from its neighbors.
Thus, each iteration can be implemented in a single round.
In the parallel setting, in each iteration, the dual variables can be raised in parallel.

The above procedure returns a pair of solutions $\calA$ and $\alpha$.
It is easy to see that $\calA$ is a feasible solution  for the given
set cover instance. Furthermore, only sets satisfying the bound (\ref{eqn:DDD})
are added to the collection $\calA$. Hence, the pair satisfies the
desired approximate primal slackness property.

Let us next analyze the number of iterations taken by the algorithm.
The number of epochs is $L$. Fix any epoch $k$.
For any iteration $j$, define the minimum penalty value $p_j^{\min} = \min_{S} p_j(S)$ 
(where the minimum is taken over all sets participating in iteration $j$). 
We now establish a bound on the number of iterations taken by the any epoch $k$,
by tracking minimum penalty value.
For a set $S$ participating in successive iterations $j$ and $j+1$,
its penalty may decrease (because both the values $\delta(S)$ and $c(S)$ may decrease across iterations).
Nevertheless, the lemma below shows that the minimum penalty will increase by a factor of at 
least $(3/2)$ across successive iterations. 

\begin{lemma}
\label{lem:RRR}
For any iteration $j$, $p_{j+1}^{\min}\geq (3/2)p_j^{\min}$.
\end{lemma}
\myproof
Let $S$ be any set participating in the $j$th iteration.
In $j$th iteration, when the dual variables are raised for the participating elements,
the LHS value of the dual constraint of $S$ will increase by some amount; let this amount be $\delta_j(S)$.
Consider the $d_j(S)$ elements contained in $S$ that participate in the $j$th iteration.
There are $d_j(S)$ elements that are uncovered by $\calA$ in the beginning of the $j$th iteration.
Of these elements, an element $e$ said to be {\em good} to $S$, if $q_j(e)\geq (1/4)p_j(e)$.
Intuitively, when we raise $\alpha(e)$ by $q_j(e)$, the LHS of the dual constraint of $S$ would raise by
at least $(1/4)p_j(S)$. We say that an element $S$ is {\em successful} in iteration $j$,
if at least $(1/2) d_j(S)$ elements are good for $S$.
As we observed earlier, the penalty of a set may decrease across iterations.
But, we next show that the penalty of an unsuccessful set cannot decrease by much.

\begin{claim}
\label{claim:EEE}
Any set $S$ successful in the $j$th iteration would be added to $\calA$ in that iteration.
\end{claim}
\myproof
Since $S$ is successful, $d_j(S)/2$ elements are good for $S$ and each would raise the LHS
value by at least $(1/4)p_j(S)$. Thus, 
\[
\delta_j(S) \geq d_j(S)p_j(e)/8 = c_j(S)/8. = (w(S)-h_j(S))/8
\]
So, after the raise in the dual variables, the LHS value will be at least $w(S)/8$.
\[
h_j(S) + \delta = (7/8)h_j(S) + w(S)/8\geq w(8)/8
\]
Therefore, $S$ will be added to $\calA$ in the $j$th iteration.
\qed

\begin{claim}
\label{claim:FFF}
Any set $S$ satisfying $p_j(S)\leq 4\cdot p_j^{\min}$ would be added to $\calA$ in that iteration.
\end{claim}
\myproof 
For such a set $S$, all the $d_j(S)$ elements will be good. Therefore, it will be successful.
\qed

\begin{claim}
\label{claim:GGG}
For any unsuccessful set $S$ that participates in the iteration $(j+1)$,
$p_{j+1}(S) \geq (3/8) p_j(S)$.
\end{claim}
\myproof
Consider the increase in LHS $\delta_j(S)$.
Since $S$ is unsuccessful, there are at most $(1/2)d_j(S)$ good elements,
each of which may contribute $p_j(S)$ towards $\delta_j(S)$.
On the other hand, the bad elements can contribute at most $(1/4)p_j(S)$.
Therefore,
\[
\delta_j(S) \leq (1/2)d_j(S)p_j(S) + (1/2)d_j(S)(1/4)p_j(S) \leq (5/8)c_j(S).
\]
It follows that
\[
c_{j+1}(S) = c_j(S) - \delta_j(S) \geq (3/8)c_j(S).
\]
Since $d_{j+1}(S) \leq d_j(S)$, we get that
\[
p_{j+1}(S) = c_{j+1}(S)/d_{j+1}(S) \geq c_{j+1}(S)/d_j(S) \geq (3/8)c_j(S)/d_j(S)=(3/8)p_j(S)
\]
\qed

Consider any set $S$ that participates in iteration $j+1$.
By Claim \ref{claim:EEE}, it must be unsuccessful. Therefore, by Claim \ref{claim:GGG},
$p_{j+1}(S) \geq (3/8) p_j(S)$. Moreover, by Claim \ref{claim:FFF}, $p_j(S) \geq 4\cdot p_j^{\min}$. 
It follows that $p_{j+1}(S) \geq (3/2)\cdot p_j^{\min}$.
We conclude that $p_{j+1}^{\min} \geq (3/2) p_j^{\min}$.
This completes the proof of the lemma.
\qed

We shall derive a bound on the number of iteration by making some observation on the maximum and minimum values possible
for $d_j(S)$ and $c_j(S)$.
The $d_j(S)$ values can vary between $1$ and $m$.
The maximum value possible for $c_j(S)$ is $w_{\max}$; 
the minimum value possible is $(7/8)w_{\min}$ 
(because sets with smaller $c_j(S)$ would have got added to $\calA$).
Therefore,  epoch $k$ will take at most 
$O(\log m + \log \frac{w_{\max}} {w_{\min}} )$ iterations.
Hence, the overall forward phase algorithm runs in 
$O(L\cdot [\log m + \log \frac{w_{\max}} {w_{\min}}])$ iterations.

We next record some useful properties satisfied by the pair of solution $\pair{\calA}{\alpha}$
output by the forward phase. These properties will be useful during the reverse-delete phase.
Partition the collection $\calA$ into $\calA_1, \calA_2, \ldots, \calA_L$, where $\calA_k$ is the
collection of sets added to $\calA$ in the epoch $k$ of the forward phase.
For $1\leq k\leq L$, let $F_k$ be the set of elements freshly covered by $\calA_k$ (meaning, the elements
covered by $\calA_k$ which are not covered by $\calA_1, \calA_2, \ldots, \calA_{k-1}$).
We say that $\calA_k$ is {\em responsible} for the elements in $F_k$.
Intuitively, in epoch $k$, the main task of the algorithm was to ensure coverage for $R_k\subseteq Z_k$
and the sets in $\calA_k$ were selected for this purpose. But some elements
belonging to $Z_{k+1}, Z_{k+2}, \ldots, Z_k$ might also be covered by $\calA_k$.
The set $F_k$ consists of $R_k$ and the above elements.

\begin{proposition}
\label{prop:XXX}
(i) For $1\leq k\leq L$, $F_k$ consists of elements only from layers $Z_k, Z_{k+1}, \ldots, Z_L$.
(ii) For $1\leq k\leq L$, the collection $\calA_k$ does not cover any element from $R_{k+1}, R_{k+2},\ldots, R_L$.
(iii) The elements found in $R_1, R_2, \ldots, R_L$ are the only elements whose dual variables could 
potentially have been raised in the forward phase.
\end{proposition}

\subsection{Reverse Delete Phase}
The forward phase produces a pair of solutions $\pair{\calA}{\alpha}$.
In the reverse delete phase, we prune the collection $\calA$ and obtain a solution $\calB\subseteq \calA$
such that the solution $\calB$ satisfies the approximate complementary slackness property: for any $e\in E$, if $\alpha(e)>0$ then
\begin{eqnarray}
\label{eqn:MMM}
|\{S\in \calB~:~S\mbox{ covers } e\}| \leq \tau^2.
\end{eqnarray}
Furthermore, we will not alter the dual variables during the reverse-delete phase.
Hence, the final pair of solutions $\calB$ and $\alpha$ satisfy 
both the primal and dual approximate complementary slackness properties, namely bounds (\ref{eqn:DDD}) and
(\ref{eqn:MMM}). The weak duality theorem implies that the solution $\calB$ is an $(8\tau^2)$-approximate solution.

We now describe the reverse-delete phase that would satisfy the bound (\ref{eqn:MMM}).
By the third part of Proposition \ref{prop:XXX},
it suffices if we consider elements in $R_1, R_2, \ldots R_L$.
The reverse delete procedure is also iterative and works in $L$ epochs, 
but it will consider the layers in the reverse direction,
namely, the iterations are from $k=L$ to $1$.
Initialize $\calB=\emptyset$. 
At the end of epoch $k$, we will ensure two properties: 
(i) all the elements in $F_L, F_{L-1}, \ldots, F_k$ are covered by $\calB$;
(ii) all the elements in $R_L, R_{L-1}, \ldots, R_k$ obey  the slackness property (\ref{eqn:MMM}).

Assume by induction that we have satisfied the above two properties in iteration $L, L-1, \ldots, k+1$
and consider epoch $k$. Our plan is to ensure coverage of $F_k$ by adding sets from $\calA_k$ to $\calB$
(recall that $\calA_k$ is responsible for $F_k$).
An important issue here is that the sets added to $\calB$ in the previous iterations 
$L, L-1, \ldots, k+1$ will be from $\calA_L, \calA_{L-1}, \ldots, \calA_{k+1}$, 
which are not responsible for covering the elements in $F_k$; nevertheless, some of these sets might still be 
covering the elements in $R_k\subseteq F_k$ (this is an unintended side-effect of the forward phase).
While ensuring slackness property (\ref{eqn:MMM}) for the elements in $R_k$,
we have to take the above phenomenon into account and may have to delete sets from $\calB$.
In doing so, we should not affect the coverage of the elements in $F_L, F_{L-1}, \ldots, F_{k+1}$.
The procedure given by the lemma below helps us in achieving the above objectives;
the lemma is proved in Section \ref{sec:XY-lemma}.

\begin{lemma}
\label{lem:HHH}
Let $A\subseteq E$ be a set of elements belonging to layers $Z_k, Z_{k+1}, \ldots, Z_L$, for some given $k$.
Let $\calX\subseteq \calS$ be a cover for $A$.
There exists a parallel procedure that takes $\calX$ and $A$ as input, and outputs a collection $\calY\subseteq \calX$
such that: (i) $\calY$ is a cover for $A$; (ii) for any element in $e\in A$ belonging to layer $Z_k$,
at most $\tau^2$ sets from $\calY$ cover $e$. 
The algorithm takes at most $L$ iterations, where the dominant operation
in each iteration is computing a maximal independent set (MIS) in an arbitrary graph.
\end{lemma}

We are now ready to discuss epoch $k$. Let $\calX = \calB \cup \calA_k$. 
Let $A=F_L\cup F_{L-1} \cup \cdots \cup  F_k$.
Notice that the requirements of the Lemma \ref{lem:HHH} are satisfied by $A$ and $\calX$
(because by induction, $\calB$ covers $F_L, F_{L-1}, \ldots, F_{k+1}$ and 
$\calA_k$ covers $F_k$).
Invoke the procedure given by the lemma and obtain a set $\calY$. 

We claim that $\calY$ satisfies two properties: 
(i) $\calY$ is a cover for $F_L, F_{L-1}, \ldots, F_k$;
(ii) for any element $e$ in $R_L, R_{L-1}, \ldots, R_k$ at most $\tau^2$ sets from $\calY$ cover $e$.
The first property is ensured by the lemma itself.
Moreover, the lemma guarantees that the second property is true for any element $e\in R_k$. 
So, consider an element $e$ belonging to one of the sets $R_L, R_{L-1}, \ldots, R_{k+1}$.
The lemma ensures that $\calY \subseteq \calX = \calB\cup \calA_k$
and hence, the sets $e$ must come from $\calB$ or $\calA_k$.
Proposition \ref{prop:XXX} implies that $\calA_k$ does not contain any set covering $e$.
Therefore, all the sets covering $e$ must come from $\calB$; by the induction hypothesis,
there are at most $\tau^2$ such sets.
We have shown that $\calB$ satisfies the induction hypothesis.
We set $\calB = \calY$ and proceed to the next epoch $k-1$.

We see that the overall algorithm produces a $8\tau^2$-approximate solution.
Let us now analyze the running time.
We can preprocess the sets so that $w_{\max}/w_{\min}$ is bounded by $m$, while incurring an increase approximation ratio by an additive factor of one (see \cite{RV}).
Computing the layer decomposition will take $O(L)$ iterations and 
the forward phase will take $O(L\log m)$ iterations, where each iteration can be implemented in NC.
The reverse delete phase consists of $L^2$ iteration, where each iteration mainly involves
computing MIS, which can be computed in NC \cite{Luby}.
Thus, when $L$ is logarithmic in $m$, the overall algorithm runs in NC and produces an $(1+8\tau^2)$-approximate solution.

\begin{figure}[htbp]
\begin{center}
\fbox{\begin{minipage}  {\textwidth} \small
\begin{tabbing}
xx\=xx\=xx\=xx\=xx\=xx\= \kill
\textbf{Begin}  \\
\> // Forward Phase: \\
\> Compute the layer decomposition $Z_1, Z_2, \ldots, Z_L$ (see Lemma \ref{lem:TTT}) \\
\> For all $e \in E$ \\
\> \> let $\calA=\emptyset$ and let $\alpha(e)=0$ \\
\> For $k=1$ to $L$ \\
\> \> let $Q = E \setminus \left( \cup_{U \in \calA} U \right)$ be the set of elements not covered by $\calA$ \\
\> \> let $R_k = Z_k \cap Q $ \\
\> \> initialize $\calA_k = \phi$ (sets selected in this epoch) \\
\> \> While $R_k \nsubseteq \cup_{U \in \calA} U$ \\
\> \> \> For each $S \in \calS \setminus \calA$ \\
\> \> \> \> let $d_j(S) = | S \cap Q |$\\
\> \> \> \> let $h_j(S) = \sum_{e\in S}\alpha(e)$ \\
\> \> \> \> let $c_j(S) = w(S) - \sum_{e\in S} \alpha(S)$ \\
\> \> \> \> let $p_j(S) = c_j(S)/d_j(S)$ \\
\> \> \> For each $e \in R_k \cap Q$ \\
\> \> \> \> $q_j(e) = \min_{(S:e\in S)} p_j(S)$ \\
\> \> \> \> Raise $\alpha(e)$ by $q_j(e)$ \\
\> \> \> For each $S \in \calS \setminus \calA$ \\
\> \> \> \> If ( $\sum_{e\in S} \alpha(e)  \geq  (1/8)\cdot w(S)$ ) \\
\> \> \> \> \> Add $S$ to $\calA$ \\
\> \> \> \> \> Add $S$ to $\calA_k$ \\
\> \> \> \> \> Recompute $Q = E \setminus \left( \cup_{U \in \calA} U \right)$ (i.e., the set of elements not covered by $\calA$) \\
\ \\
\> // Reverse-delete Phase: \\
\> Initialize $\calB = \phi$ \\
\> For $k=L$ down to $1$ \\
\> \> let $F_k = \calA_k \setminus ( \cup_{i=1}^{k-1} \calA_i )$ \\
\> \> let $\calX = \calB \cup \calA_k$ \\
\> \> let $A=F_L\cup F_{L-1}, \cup F_{k+1}\cup F_k$ \\
\> \> initialize $\calY=\emptyset$ \\
\> \> partition the set $A$ according to the layers: for $k\leq j\leq L$, let $A_j=A\cap Z_j$ \\
\> \> For $j=k$ to $L$ \\
\> \> \> Let $\wt{A}_j$ be the elements of $A_j$ not covered by $\calY$ \\
\> \> \> Construct a graph $G_j$ with $\wt{A}_j$ as the vertex set; \\
\> \> \> \> add an edge between two vertices $e_1,e_2\in \wt{A}_j$ if $e_1,e_2 \in S$ for some $S\in \calX$ \\
\> \> \> Find an MIS $B_j$ within the graph $G_j$ \\
\> \> \> For each $e\in B_j$ add its petals to the collection $\calY$ \\
\> \> update $\calB = \calY$ \\
\> Output $\calB$ \\
\textbf{End}
\end{tabbing}
\end{minipage}
} 
\end{center}
\caption{Parallel Algorithm}
\label{fig:par}
\end{figure}

\subsection{Proof of Lemma \ref{lem:HHH}}
\label{sec:XY-lemma}
We initialize $\calY=\emptyset$. Partition the set $A$ according to the layers: for $k\leq j\leq L$, let $A_j=A\cap Z_j$. 
We process the sequence $A_k, A_{k+1}, \ldots, A_L$ iteratively -- in each iteration $j$,
we will add some appropriate sets from $\calX$ to $\calY$ so as to ensure coverage for all elements in $A_j$.

Consider any element $e\in A$. Let $A_j\subseteq Z_j$ be the partition to which $e$ belongs.
Let $\calP(e)\subseteq \calX$ be the collection of all sets found in $\calX$ which contain $e$.
By the properties of layered decompositions, $e$ is a $\tau$-SNC element within $Z_j\cup Z_{j+1}\cup \cdots \cup Z_L$.
Hence, there exist sets 
$S_1, S_2, \ldots, S_r\in \calP(e)$ (with $r\leq \tau$) such that
any element $e\in Z_j\cup Z_{j+1}\cup \cdots Z_L$ covered by $\calP(e)$
is also covered by one of $S_1, S_2, \ldots, S_r$.
We call these $r$ sets as the {\em petals} of $e$.

For $j=k$ to $L$, iteration $j$ is described next.
Of the elements in $A_j$, some of the elements would already be covered by $\calY$.
Let the set of remaining uncovered elements be $\wt{A}_j$.
Construct a graph $G_j$ with $\wt{A}_j$ as the vertex set; add an edge between two vertices $e_1,e_2\in \wt{A}_j$,
if some set $S\in \calX$ includes both of them. Find an MIS $B_j$ within the graph $G_j$. 
We call the elements in $B_j$ as {\em anchors}.
For each anchor $e\in B_j$ add its petals to the collection $\calY$.
Proceed to the next iteration.

We now prove that the collection $\calY$ constructed by the above process satisfies the
properties stated in the lemma. First, consider the coverage property.
For $k\leq j\leq L$, let us argue that $\calY$ covers $A_j$.
In the beginning of iteration $j$, $\calY$ would have already covered some elements
from $A_j$. So, we need to bother only about the remaining elements $\wt{A}_j$.
Consider any element $e\in \wt{A}_j$. If $e$ was selected as part of the MIS $B_j$,
then $e$ is covered by its petals. Otherwise, there must exist some element $a\in B_j$
such that $e$ and $a$ share an edge in $G_j$. This means that some set $S\in \calX$
contains both $e$ and $a$. Therefore one of the petals of $a$ would cover $e$.
Since we added all the petals of $a$ to $\calY$, $\calY$ would cover $e$.

Consider the second part of the lemma. 
We shall first argue that any two anchors are independent: namely, for any two anchors,
$a_1$ and $a_2$, no set $S\in \calX$ contains both of them.
By contradiction, suppose some set $S\in \calX$ contains both $a_1$ and $a_2$.
Consider two cases: (i) the two elements belong to the same layer;
(ii) they belong to different layers.
The first case will contradict the fact that $B_j$ is an MIS, where $j$ is the layer to which both the anchors belong.
For the second case, suppose $a_1\in A_{j_1}$ and $a_2\in A_{j_2}$ with $j_1\leq j_2$.
Our assumption is that the set $S$ contains both $a_1$ and $a_2$.
This would mean that $a_2$ will belong to one of the petals of $a_1$.
Hence, in the beginning of the iteration $j_2$, the collection $\calY$ would have already covered $a_2$.
This contradicts the fact that $a_2$ is an anchor.

We return to the second part of the lemma.
Consider any element $e\in A_k$. We analyze two cases: (i) $e$ is an anchor;
(ii) $e$ is not an anchor. In the first case, since the anchors are independent,
the petals of no other anchor can include $e$. So, the only sets in $\calY$
which include $e$ are the petals of $e$ itself; the number of such petals is at most $\tau$.
Now, consider the second case. Let $C$ be the set of all anchors $a$ such that
at least one petal of $a$ includes $e$. We claim that $|C|\leq \tau$.
By contradiction, suppose $|C|\geq \tau+1$.
Take any $\tau+1$ anchors $a_1, a_2, \ldots, a_{\tau+1}$ found in $C$.
The element $e$ belongs to the layer $Z_k$. So, it will be a $\tau$-SNC element within
$Z_k\cup Z_{k+1}\cup \cdots Z_L$. 
Hence, the petals of $e$ will cover all the anchors $a_1, a_2, \ldots, a_{\tau+1}$.
But, the number of petals of $e$ is at most $\tau$.
Hence, by the pigeon hole principle, two of these anchors must be covered by the 
same petal of $e$. This contradicts our previous claim that the anchors are independent.
Therefore, $|C|\leq \tau$. The element may belong to more than one petal of an anchor.
Each anchor $a_i\in C$ has at most $\tau$ petals. 
It follows that at most $\tau^2$ petals of the anchors can cover $e$.
This proves the second part of the claim.

\section{A Distributed Algorithm for the $\tau$-SNC Set Systems}
In this section, we describe a distributed algorithm for the set cover problem on $\tau$-SNC set systems
having an approximation ratio of $\tau$. It runs in $O(\log^2 m + L \log m)$
communication rounds, where $L$ is the decomposition length.
Thus when $L$ is logarithmic in $m$, the number of rounds in bounded by $O(\log^2 m)$.
The algorithm is obtained by implementing the sequential algorithm in a distributed fashion by appealing to the Linial-Saks decomposition \cite{LS}.

The Linial-Saks decomposition goes via the notion of color class decompositions, described next.
Let $G=(U,H)$ be a graph. A {\em color class decomposition}
of the graph $G$ is a partitioning the vertex set $U$ into {\em clusters}
$U_1, U_2, \ldots, U_r$.
The decomposition also specifies a set of {\em color classes}
$\{\calC_1, \calC_2, \ldots, \calC_d\}$
and places each cluster $U_i$ in exactly one of the color classes.
The decomposition must satisfy the following property:
any two clusters $C_i$ and $C_j$ placed in the same color class
must be independent; meaning, there should not be an edge in $H$
connecting some vertex $u\in C_i$ with some vertex $v\in C_j$.
We shall measure the efficacy of the decomposition using two parameters:
\begin{itemize}
\item
{\em Diameter}: For a cluster $C_i$, let $\ell_i$ be the maximum distance (number of hops in the
shortest path) between any pair of vertices in $C_i$. Then, the diameter of the decomposition
is the maximum of $\ell_i$ over all the clusters.
\item
{\em Depth}: The depth of the decomposition is the number of color classes $d$.
\end{itemize}

Linial and Saks \cite{LS} showed that any graph has
a decomposition with $O(\log m)$ diameter and $O(\log m)$ depth, where $m$ is the number of vertices in the graph.
They also presented a randomized distributed algorithm for finding such a decomposition
running in $O(\log^2 m)$ communication rounds.

We now describe the distributed algorithm. 
Let $\pair{E}{\calS}$ be the given set system.
The first step is to compute the the Linial-Saks decomposition of the graph determined by the communication network
of the set system.
Let $U_1, U_2, \ldots, U_r$ be the clusters and 
$\calC_1, \calC_2, \ldots, \calC_d$ be the color classes, where the depth $d=O(\log m)$.
For each cluster $U_i$, we select a leader (say the element having the least ID).
Since the diameter of the cluster is $O(\log m)$, the leader can collect all the input data
known to the elements in the cluster in a single communication round.
The leader of the cluster will do all the processing for a cluster.

Compute the layer decomposition of the given set system $Z_1, Z_2, \ldots, Z_L$ (see Lemma \ref{lem:TTT}).
The algorithm consists of a forward phase and reverse-delete phase.
We first describe the forward phase procedure which will process the layers from left to right.
It runs in $L$ epochs, where epoch $k$ will process the layer $Z_k$, as follows.
We take a pass over the color classes $\calC_1, \calC_2,\ldots, \calC_d$ in $d$ steps,
where step $j$ will handle the color class $\calC_j$ and process each cluster in the color class $\calC_j$.
For a cluster $U_i\in \calC_j$, the leader will consider all the elements in the
belonging to the layer $Z_k$ raise their dual variables using the same mechanism used in the
sequential algorithm. For each element adjacent to the some element in the cluster, 
the leader will then communicate the new values of the relevant dual variables and newly selected sets.
Since the clusters in any color class are independent,
the clusters of a color class can be processed simultaneously.
Each step can be implemented in $O(1)$ communication rounds.
The reverse-delete phase is similar, but processes the elements in the reverse order
and simulates the sequential algorithm.
The pseudo-code is presented in Figure \ref{fig:dist}.
The algorithm will run in $O(L\cdot d)$ communication rounds.
Since the construction of the Linial-Saks decomposition takes $O(\log^2 m)$ rounds,
the overall algorithm runs in $O(\log^2 m + L\log m)$ communication rounds.

\begin{figure}
\begin{center}
\begin{boxedminipage}{\hsize}
\begin{small}
\begin{tabbing}
xx\=xx\=xx\=xx\=xx\=xx\=xx\=xx\=xx\=xx\kill
\textbf{Begin}  \\
\> // Forward Phase:\\
\> Initialize. $\calA \leftarrow \emptyset$. For all $e\in E$, $\alpha(e)=0$.\\
\> For $i=1$ to $L$\\
\> \> For $j=1$ to $d$\\
\> \> \> For each cluster $U_i$ in the color class $\calC_j$\\
\> \> \> \> Let $R$ be the elements in $U_i$ belonging to layer $Z_k$.\\
\> \> \> \> Arrange the elements in $R$ in some arbitrary order $\sigma_{i,k}$.\\
\> \> \> \> For each element $e$ in $\sigma_{i,k}$\\
\> \> \> \> \> If $e$ is not covered by $\calA$\\
\> \> \> \> \> \> Raise the dual variable $\alpha(e_k)$ until some dual constraint becomes tight:\\
\> \> \> \> \> \> \> $\alpha(e_k) \leftarrow \max_{S:e_k\in S} w(S) - \sum_{a\in S} \alpha(a)$\\
\> \> \> \> \> \> \> Include the corresponding set $S$ in $\calA$:\\
\>\\
\> // Reverse-delete phase:\\
\> $\calB = \emptyset$\\
\> For $i=L$ to $1$\\
\> \> For $j=1$ to $d$\\
\> \> \> For each cluster $U_i$ in the color class $\calC_j$\\
\> \> \> \> Scan the ordering $\sigma_{i,k}$ in the reverse order.\\
\> \> \> \> For each element $e$, if $\alpha(e)$ was raised in the forward phase do:\\
\> \> \> \> \> Let $X$ be the neighbors of $e$ not covered by $\calA$ when $\alpha(e)$ was raised.\\
\> \> \> \> \> Let $\calP\subseteq \calB$ be the collection of sets covering $e_k$\\
\> \> \> \> \> Find sets $S_1, S_2, \ldots, S_l\in \calP$ (with $l\leq \tau$) such that\\
\> \> \> \> \> \> all the elements in $X$ covered by $\calP$ are also covered by $S_1, S_2, \ldots, S_r$\\
\> \> \> \> \> Delete all the sets found in $\calP$ from $\calB$, except $S_1, S_2, \ldots, S_r$\\
\> Output $\calB$.\\
\textbf{End}
\end{tabbing}
\end{small}
\end{boxedminipage}
\end{center}
\caption{Distributed Algorithm}
\label{fig:dist}
\end{figure}

\section{Computing Layer Decomposition : Proof of Lemma \ref{lem:TTT}}
\label{sec:layer-compute}
We first present a polynomial time procedure
that take as input subset of elements $X\subseteq E$ and an element $e\in X$,
and tests whether $e$ is a $\tau$-SNC element within $X$.

The following notation is useful in this context.
Let $\pair{E}{\calS}$ be the input set system.
Let $\calQ \subseteq \calS$ be the collection of all sets that include $e$.
We say that a subset $\calP \subseteq \calQ$ is $\tau$-collapsible, if
there exist $\tau$ sets $S_1, S_2, \ldots, S_r\in \calP$ such that
every element in $X$ covered by $\calP$ is also covered by one of the above $\tau$ sets
and $r\leq \tau$; the $\tau$ sets are called the {\em base sets} of $\calP$.
Testing whether $e$ is a $\tau$-SNC element within $X$ is the same as testing
whether every collection $\calP\subseteq \calQ$ is $\tau$-collapsible.
A naive algorithm would enumerate all the possible subsets of $\calQ$ and test whether each one of them
is $\tau$-collapsible. However, such an approach may take exponential time.
The following combinatorial lemma helps in obtaining a polynomial time procedure.

\begin{lemma}
\label{lem:SSS}
Suppose every collection $\calA \subseteq \calQ$ of cardinality $\tau+1$ is $\tau$-collapsible.
Then, every collection $\calP\subseteq \calQ$ is $\tau$-collapsible.
\end{lemma}
\myproof
Consider any collection $\calP\subseteq \calQ$ having cardinality at least $\tau+1$ 
(the claim is trivially true for smaller collections). 
Let the sets contained in the collection be $P_1, P_2, \ldots, P_s$ (for some $s\geq \tau+1$), arranged
in an arbitrary manner.
Via induction, we shall argue that for any $k\geq \tau+1$, the collection 
$\{P_1, P_2, \ldots, P_k\}$ is $\tau$-collapsible.
For the base case, the collection 
$\{P_1, P_2, \ldots, P_{\tau+1}\}$; this collection is $\tau$-collapsible by the hypothesis of the lemma.
By induction, suppose the claim is true for the collection $\{P_1, P_2, \ldots, P_k\}$.
Now, consider the collection $\{P_1, P_2, \ldots, P_{k+1}\}$.
If $r<\tau$, the we can simply add $P_{k+1}$ to the sequence $S_1, S_2, \ldots, S_r$
and get the base sets for the above collection. So, assume that $r=\tau$.
By our hypothesis, the collection $\{S_1, S_2, \ldots, S_r, P_{k+1}\}$ must be $\tau$-collapsible.
Let the collection of base sets of for the above collection be $S_1', S_2', \ldots, S_q'$.
Observe that $S_1', S_2', \ldots, S_q'$ form base sets for the collection $\{P_1, P_2, \ldots, P_{k+1}\}$.
Thus, we have proved the claim. The lemma follows by taking $k=s$.
\qed

Based on the above lemma, it suffices if we consider collections $\calA\subseteq \calQ$ of cardinality $\tau+1$.
The number of such collections is at most $m^{O(\tau)}$, where $m=|E|$. 
For each such collection, we can test $\tau$-collapsibility 
in time polynomial in $m$.
Since $\tau$ is assumed to be a constant, 
this yields a polynomial time procedure for testing an element $e$ is a $\tau$-SNC element within a set $X$.

It is now easy to compute the layer decomposition of the given set system $\pair{E}{\calS}$.
We consider every element $e\in E$ and test whether $e$ is a $\tau$-SNC element within $E$.
All the elements passing the test are placed in $Z_1$.
We remove these elements and apply the same procedure on the remaining set of elements.
After $L$ iterations, we would have computed the layer decomposition.

The above procedure runs in polynomial time in the sequential setting.
In the distributed setting, the algorithm can be implemented in $O(L)$ communication rounds.
In the parallel setting, each of the $L$ iterations can be implemented in NC.

\section{Bound on the Decomposition Length for the Tree Cover Problem}
\label{sec:tree}
Recall that the set systems induced by the tree cover problem satisfy the $1$-SNC property 
with decomposition length bounded by the depth of the tree. 
Such a layer decomposition would not be sufficient for obtaining polylogarithmic time bounds.
In this section we show that the set systems induced by the tree cover problem are $2$-SNC set systems
having decomposition length only $\log m$.

In the given tree $T$, we say that a node $z$ is a {\em junction}, if it has more than one children nodes.
It will be convenient to consider the root also as a junction, even if it has only one child.
Consider any leaf node $v$. Let $p$ be the path connecting the root and $v$. 
Starting from the node $v$ traverse up the path $p$
until we hit a junction $z$ (or the root node itself). Consider the path $q$ connecting $z$ and $v$;
we call $q$ as the {\em chain} defined by the leaf node $v$ in the tree $T$.
Let $e$ be any edge on the path $q$. We claim that $e$ is a $2$-SNC element.
Consider any set of interval $\calP$ covering the edge $e$.
Among these intervals, let $I_l$ be the interval extending the most towards the leaf node $v$
and let $I_r$ be the interval extending the most towards the root node.
Notice that for any interval $I\in \calP$, the intervals $I_l$ and $I_r$ put together cover
all the edges covered by $I$. This shows that all the edges found on the chain $q$ are $2$-SNC elements.
In general, let $\{v_1, v_2, \ldots, v_r\}$ be the set of all leaf nodes in $T$.
Let $q_1, q_2, \ldots, q_r$ be the chains defined by the above leaf nodes.
Then, all the edges found along these chains will be $2$-SNC elements.

We shall apply the above procedure iteratively to decompose the set of all edges into chains.
Let $T_1=T$ be the given tree. Consider iteration $k\geq 1$.
Find all the leaf nodes in the tree $T_k$. Compute the chains defined by these leaf nodes.
Create a group $B_k$ and put all the edges found on these chains in the group $B_k$.
Delete all these edges along with their vertices, except for the junctions.
Let the remaining tree be $T_{k+1}$. We then proceed to the iteration $k+1$,
and process the tree $T_{k+1}$. We terminate the process when there are no more edges left.
The iterative procedure will terminate after some $K$ iterations,
yielding groups $B_1, B_2, \ldots, B_K$. We call $B_1, B_2, \ldots, B_K$
as the {\em chain decomposition} of the given tree $T$. The quantity $K$ is called the
{\em length} of the above decomposition. 

Let $Z_1, Z_2, \ldots, Z_L$ be the $2$-SNC layer decomposition of the set system.
We next prove that $L\leq K$. We argued that all the edges in $B_1$ are $2$-SNC elements 
within the entire universe $E=H$.
Extending this argument, we can show that for $k\geq 1$, the edges in $B_k$ will be $2$-SNC elements within $E_k$,
where $E_k$ is the set of edges in the tree $T_k$.
(Intuitively, this means that the edges in $B_k$ will belong to layer $k$. 
However, it is possible that some edges from $B_k$ may belong to a lower layer;
this depends on how the input intervals are constructed). 
Using the above fact, we can formally show that for any $k\leq 1$,
any edge $e\in B_k$ is found in some layer $j\leq k$ (i.e., $e\in Z_j$).
It follows that $L\leq K$.

Our next task is to prove a bound on $K$.
Consider the sequence of trees $T_1, T_2, \ldots, T_K$.
Let $\ell_1, \ell_2, \ldots, \ell_K$ be the number of leaf nodes in these trees, respectively.
We claim that for $1\leq k \leq K-1$, $\ell_{k+1}\leq \ell_k/2$.
To see this, first notice that the leaf nodes of $T_{k+1}$ are exactly the junctions in $T_k$.
Thus, $J_k = \ell_{k+1}$, where $J_k$ are the number of junctions in $T_k$.
Each junction in $T_k$, by definition, would have at least $2$ leaf nodes in the sub-tree beneath it.
Hence, $\ell_k \geq 2\cdot J_k$. Thus the claim is proved. It follows that the number of leaf nodes
reduces by a factor of at least two in each iteration. Hence, $K$ is at most $\log m$
and therefore, $L\leq \log m$.

\section{Conclusions and Open Problems}
In this paper, we introduced the concept of $\tau$-SNC set systems
and presented a sequential $\tau$-approximation algorithm for the set cover problems on such systems.
For the case where the decomposition length is logarithmic, we presented distributed 
and parallel algorithms with approximation ratios of $\tau$ and $(1+8\tau^2)$, respectively.
The parallel algorithm raises the following interesting open questions:
(i) In the forwards phase, can a $(1-\epsilon)$-maximal dual solution
be produced in number of iterations independent of $f$? 
(ii) The reverse delete phase, the algorithm prodcues a primal intergal solution satisfying the primal slackness property
with parameter $\tau^2$. Can this be improved to $\tau$? (iii) The zig-zag nature of the reverse delete phase
leads to $L^2$ iterations. Can this be improved to $L$?
Both the distributed and parallel algorithms take number of rounds dependant on $L$.
If this dependence can be removed, then we can hope to construct constant factor approximation
algorithms for $\tau$-SNC set cover problems of arbitrary decomposition length (rather than logarithmic decomposition length 
addressed in the current paper).

\bibliographystyle{plain}
\bibliography{main}

\end{document}